\documentclass[aps,prl,superscriptaddress,showpacs,twocolumn]{revtex4-1}
\usepackage{graphicx}
\usepackage{amsmath,amsfonts}
\usepackage{dsfont}
\usepackage{xspace}
\usepackage{color}
\usepackage[colorlinks=true,linkcolor=blue,urlcolor=black,citecolor=blue]{hyperref}
\usepackage{stmaryrd}
\usepackage{pdfpages}

\DeclareMathOperator{\re}{\mbox{Re}}

\newcommand{\eqw}[1]{(\ref{#1})}
\newcommand{\eq}[1]{Eq.\thinspace{}(\ref{#1})}

\newcommand{\fig}[1]{Fig.\thinspace{}\ref{#1}}

\newcommand{\fc}[1]{({#1})}

\def\ket#1{\mathinner{|{#1}\rangle}}

\makeatletter
\newsavebox{\@brx}
\newcommand{\llangle}[1][]{\savebox{\@brx}{\(\m@th{#1\langle}\)}%
  \mathopen{\copy\@brx\kern-0.5\wd\@brx\usebox{\@brx}}}
\newcommand{\rrangle}[1][]{\savebox{\@brx}{\(\m@th{#1\rangle}\)}%
  \mathclose{\copy\@brx\kern-0.5\wd\@brx\usebox{\@brx}}}
\makeatother

\def\beq{\begin{equation}}
\def\eeq{\end{equation}}
\def\bea{\begin{eqnarray}}
\def\eea{\end{eqnarray}}

\begin{document}
\title{Interferometric probes of many-body localization}

\author{M. Serbyn}
\thanks{M.S., M.K., and S.G. contributed equally to this work.}
\affiliation{Department of Physics, Massachusetts Institute of Technology, Cambridge MA 02139, USA}

\author{M. Knap}
\thanks{M.S., M.K., and S.G. contributed equally to this work.}
\affiliation{Department of Physics, Harvard University, Cambridge MA 02138, USA}
\affiliation{ITAMP, Harvard-Smithsonian Center for Astrophysics, Cambridge MA 02138, USA}

\author{S. Gopalakrishnan}
\thanks{M.S., M.K., and S.G. contributed equally to this work.}
\affiliation{Department of Physics, Harvard University, Cambridge MA 02138, USA}

\author{Z. Papi\'c}
\affiliation{Perimeter Institute for Theoretical Physics, Waterloo, N2L2Y5 ON, Canada}
\affiliation{Institute for Quantum Computing, Waterloo, Ontario N2L 3G1, Canada}

\author{N. Y. Yao}
\affiliation{Department of Physics, Harvard University, Cambridge MA 02138, USA}

\author{C. R. Laumann}
\affiliation{Department of Physics, Harvard University, Cambridge MA 02138, USA}
\affiliation{Perimeter Institute for Theoretical Physics, Waterloo, N2L2Y5 ON, Canada}
\affiliation{Department of Physics, University of Washington, Seattle WA 98195, USA}

\author{D. A. Abanin}
\affiliation{Perimeter Institute for Theoretical Physics, Waterloo, N2L2Y5 ON, Canada}
\affiliation{Institute for Quantum Computing, Waterloo, Ontario N2L 3G1, Canada}

\author{M. D. Lukin}
\affiliation{Department of Physics, Harvard University, Cambridge MA 02138, USA}

\author{E. A. Demler}
\affiliation{Department of Physics, Harvard University, Cambridge MA 02138, USA}

\date{\today}

\begin{abstract}

We propose a  method for detecting many-body localization (MBL) in disordered spin systems. The method involves pulsed, coherent spin manipulations that probe the dephasing of a given spin due to its entanglement with a set of distant spins. It allows one to distinguish the MBL phase from a non-interacting localized phase and a delocalized phase. In particular, we show that for a properly chosen pulse sequence the MBL phase exhibits a characteristic power-law decay reflecting its slow growth of entanglement.  We find that this power-law decay is robust with respect to thermal and disorder averaging, provide numerical simulations supporting our results, and discuss possible experimental realizations in solid-state and cold atom systems.
 
\end{abstract}

\pacs{
75.10.Jm %Quantized spin models, including quantum spin frustration 
05.70.Ln %Nonequilibrium and irreversible thermodynamics 
72.15.Rn %Localization effects (Anderson or weak localization) 
}

\maketitle

{\it Introduction.---}One of the central assumptions of statistical mechanics, which underlies conventional kinetic and transport theories, is that interactions between particles establish local equilibrium. 
This assumption, however, was recently shown to fail in a class of disordered interacting systems~\cite{basko_metalinsulator_2006,gornyi_interacting_2005,oganesyan_localization_2007,znidaric_many-body_2008, monthus_many-body_2010,pal_many-body_2010, cuevas_level_2012,bardarson_unbounded_2012,vosk_many-body_2013,iyer_many-body_2013,serbyn_universal_2013,serbyn_local_2013,huse_phenomenology_2013,huse_localization_2013, bauer_area_2013,swingle_simple_2013,pekker_hilbert-glass_2013,vosk_dynamical_2013,bahri_localization_2013,
chandran_many-body_2013,mueller_2013,rahul_coupling_2013}. Strong enough disorder can give rise to a {\it many-body localized} (MBL) phase, in which transport is absent and the system cannot act as a heat bath for its constituent parts. 
Although the MBL phase resembles a conventional, noninteracting Anderson insulator in that diffusion is absent, it has very different dynamical properties.  Specifically, interactions between particles in the MBL phase can cause dephasing and generate long-range entanglement, leading to the slow growth of entanglement entropy~\cite{bardarson_unbounded_2012, vosk_many-body_2013,serbyn_universal_2013, serbyn_local_2013, huse_phenomenology_2013}. 

\begin{figure}[bh!]
\begin{center}
 \includegraphics[width=0.49\textwidth]{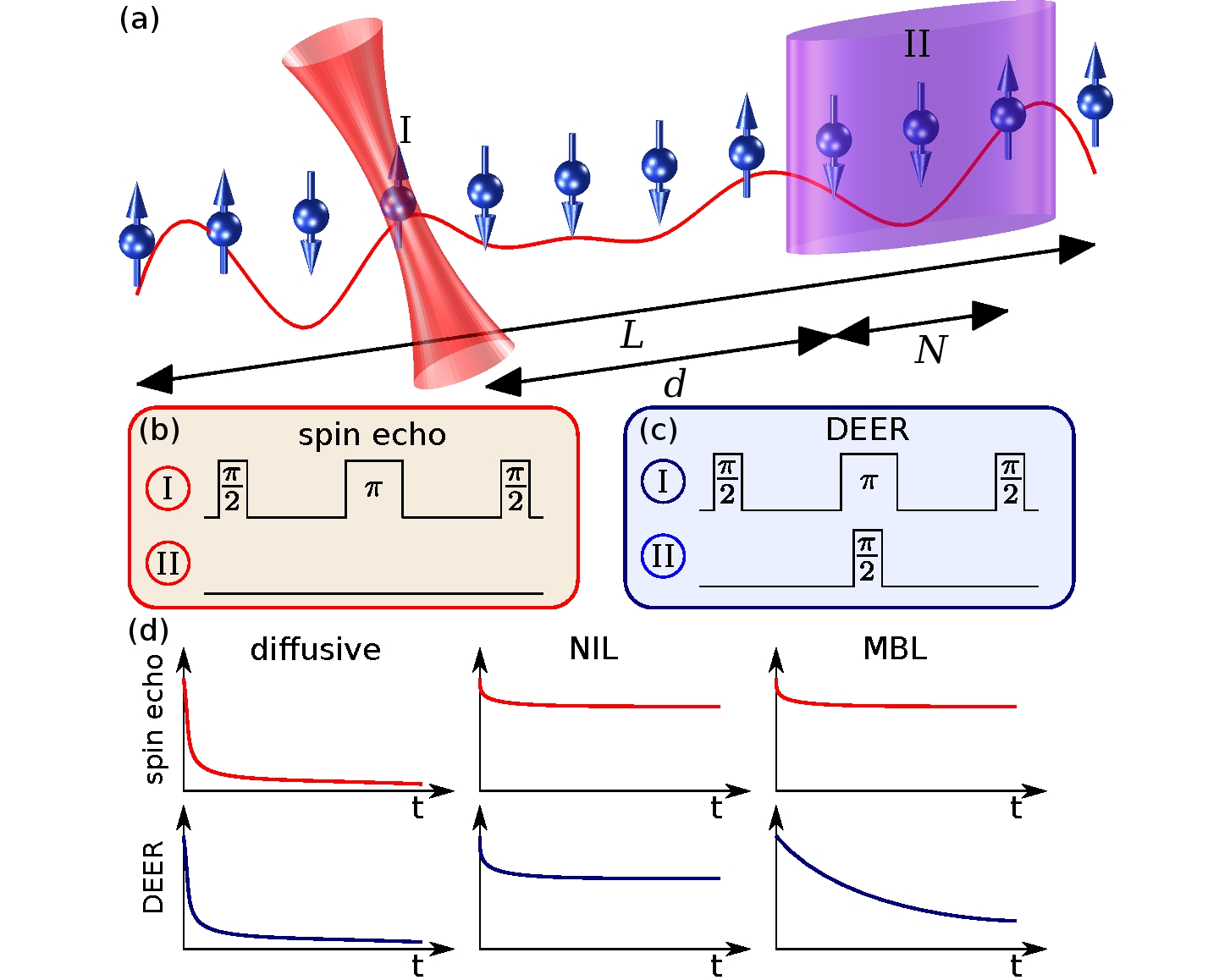}
\end{center}
\caption{\label{fig:schematic} (Color online) Schematic illustration of the proposed protocols. 
\fc{a} Spins are manipulated with lasers in two spatially separated regions I and II. \fc{b} A Hahn spin-echo sequence is applied to region I while leaving region II untouched. \fc{c} The DEER protocol differs by $\pi/2$ rotations in region II which are performed after half of the evolution time.
\fc{d} Schematic response of a system in the diffusive (left), non-interacting localized (center), and many-body localized (right) phases, to spin-echo and DEER protocols respectively. The combined information from both sequences allows one to distinguish the different phases.
}
\end{figure}

Experimental investigations of MBL in conventional solid-state systems~\cite{ovadyahu} are challenging, as these systems are strongly coupled to the environment~\cite{rahul_coupling_2013}, e.g., due to the presence of phonons. However, recent experimental advances have resulted in the realization of isolated, synthetic many-body systems with tunable 
interactions and disorder, which constitute promising platforms to explore MBL. Such systems include ultracold atoms in optical lattices~\cite{demarco1, Inguscio, demarco2}, polar molecules~\cite{polar,yan_observation_2013}, and isolated spin impurities in solids~\cite{NV1,NV2}. Although conventional transport experiments are challenging in these systems, they often allow for the precise manipulation of individual degrees of freedom to characterize their quantum evolution. This  motivates the development of new approaches for detecting and exploring the MBL phase. 

In this paper, we propose and analyze a new method for studying MBL, based on coherent manipulation of individual degrees of freedom. We focus on disordered spin systems, and show that spin-echo type measurements performed on individual spins can be used as sensitive probes of localization [Fig.~\ref{fig:schematic}(a)-(c)]. Such measurements are standard in bulk liquid and solid-state spin systems (see~\cite{vandersypen_nmr_2005} and references therein), 
and have recently been extended to probe many-body physics~\cite{bloch_echo_2003, stamper_spin_echo_2011,knap_time-dependent_2012,knap_probing_2013, bahri_localization_2013}. 
 
 Specifically, in order to probe MBL, we introduce a modified, non-local spin-echo protocol [Fig.~\ref{fig:schematic}(c)],  akin to the double electron-electron resonance (DEER) technique in electron spin resonance~\cite{wang_nmr_1984, milov_electron-electron_1984, larsen_double_1993}, that allows one to probe the dynamical correlations between remote regions of a many-body system. This approach can reveal 
interaction effects and probe quantum entanglement within the MBL phase. In particular,  the slow growth of entanglement entropy associated with the MBL phase manifests itself in a power-law decay of the DEER response. Furthermore, by measuring both the spin-echo and DEER response one can distinguish the MBL phase from a non-interacting localized (NIL) phase as well as a diffusive phase [Fig.~\ref{fig:schematic}(d)]. We discuss specific realizations of our proposal in several cold atom and solid state systems. 

{\it Approach.---} The key idea of this work can be illustrated using a phenomenological model of the MBL phase~\cite{huse_phenomenology_2013, serbyn_local_2013} that characterizes it by an infinite number of \emph{local} integrals of motion, which can be chosen as effective spin-1/2 operators $\tau_i^z$ with eigenvalues $\pm 1$.
In terms of these variables, the MBL Hamiltonian is~\cite{huse_phenomenology_2013, serbyn_local_2013}
\beq\label{simpleH}
\hat H = \sum_i \tilde{h}_i \tau^z_i + \sum_{ij} \mathcal{J}_{ij} \tau^z_i \tau^z_j+ \sum_{ijk} \mathcal{J}_{ijk} \tau^z_i \tau^z_j \tau^z_k +\;\ldots.
\eeq
The couplings $\mathcal{J}_{ij}, \mathcal{J}_{ijk}, \ldots$ fall off exponentially with separation with a characteristic localization length $\xi$ (expressed in units of the lattice constant). The Hamiltonian (\ref{simpleH}) conserves the expectation value of each $\tau_i^z$; however, interactions between effective spins randomize relative phases of different components of the wave function. Such dephasing generates entanglement between distant parts of the system~\cite{serbyn_local_2013, huse_phenomenology_2013}.

We first discuss interferometric signatures of Hamiltonian~\eqref{simpleH} assuming that one can directly manipulate the effective spins $\tau_i^z$  (in what follows we shall refer to effective spins simply as ``spins''), and later generalize these arguments to realistic cases involving manipulation of \emph{physical} rather than effective spins.

Let us first consider a simple spin-echo sequence applied to an individual spin I [Fig.~\ref{fig:schematic}(b)]. Starting from an arbitrary eigenstate of $\hat H$ (i.e., a product state of the form $\ket{\uparrow \downarrow \downarrow \uparrow \downarrow \ldots }$), we initialize spin I in a superposition state $\ket{+}_\text{I} = (\ket{\uparrow}_\text{I}  + \ket{\downarrow}_\text{I} )/\sqrt{2}$. Spin I precesses in the magnetic field $h_\text{eff}(\text{I}) = \tilde{h}_\text{I} + \sum_{j}\mathcal{J}_{\text{I}j} \tau^z_j+ \sum_{j,k}\mathcal{J}_{\text{I}jk}  \tau^z_j\tau^z_k + \ldots$, which depends on the state of the surrounding spins. The thermal average over initial states gives rise to dephasing and decay of the free precession signal. The standard spin-echo sequence, however, allows one to recover the quantum coherence of spin I, by applying a time-reversal $\pi$-pulse to it at time $t/2$. For the MBL Hamiltonian~\eqref{simpleH}, the precession induced by $h_\text{eff}(\text{I})$ over the 
initial 
evolution for $t/2$ is cancelled by the precession accumulated during evolution for time $t/2$ after the $\pi$-pulse, independent of the value of $h_\text{eff}(\text{I})$. However, since spin echo is insensitive to dephasing in the MBL phase, it does not distinguish between NIL and MBL phases. 

\begin{figure}[t] 
\begin{center}
\includegraphics[width=0.99\columnwidth]{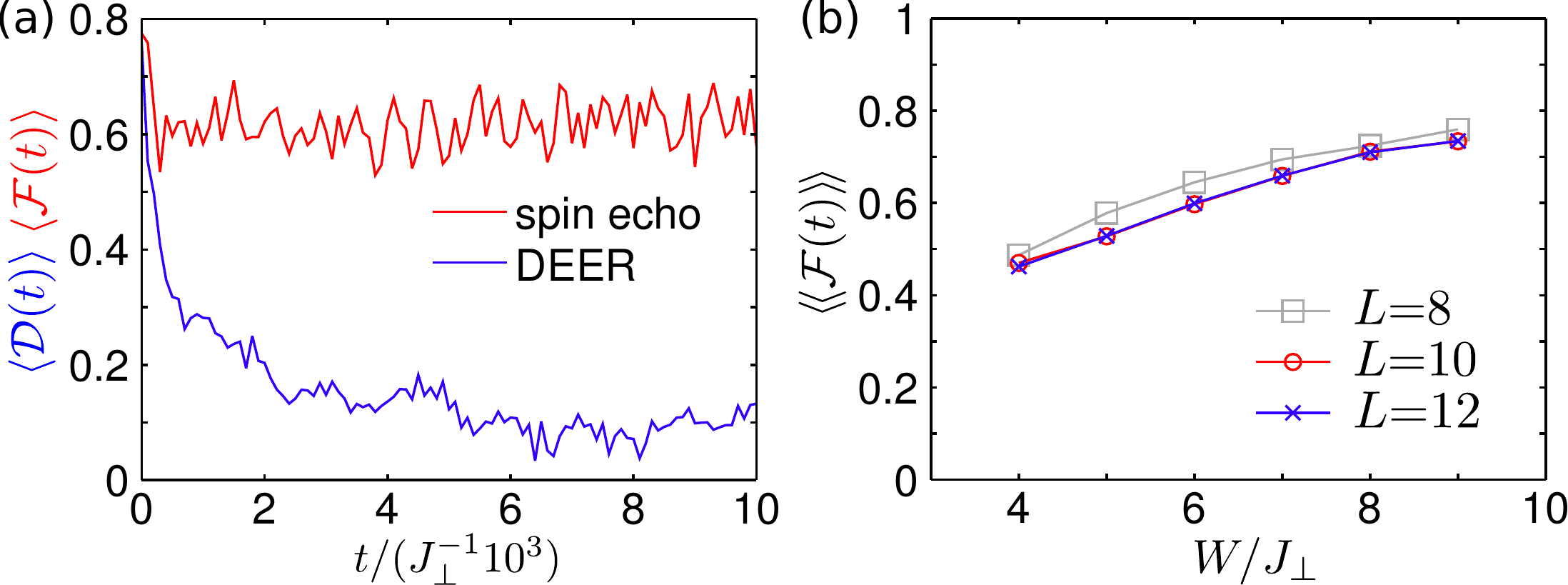}
\caption{ (a) Typical behavior of spin-echo and DEER response for the random-field XXZ model [Eq.~\eqref{eq:h}], averaged over 50 random eigenstates for  a single disorder realization ($W=6$). The spin-echo response $\mathcal{F}(t)$ quickly saturates, whereas the DEER response $\mathcal{D}(t)$ slowly decays to a much smaller value. (b) Saturation values of \emph{disorder-averaged} spin-echo fidelity as a function of disorder strength $W$ and system size $L$ for the random-field XXZ model with $J_z = J_\perp$. These results are consistent with the expectation  that $\llangle \mathcal{F}(t) \rrangle$ should saturate to a nonzero value in the thermodynamic limit (see main text).}
\label{fig2}
\end{center}
\end{figure}

We next  introduce a modified spin-echo protocol, which directly probes interaction effects in the MBL phase. The idea, inspired by the DEER technique~\cite{wang_nmr_1984,milov_electron-electron_1984, larsen_double_1993}, is to perturb spins in a remote region II, situated at a distance $d\gtrsim \xi$ away from I, halfway through the spin-echo sequence. More specifically, DEER is identical to spin echo for the first $t/2$ of the time evolution, but simultaneously with the $\pi$ pulse to spin $\text{I}$, another pulse (which we shall take to be a $\pi/2$ pulse) is applied to all the spins in region II. Assuming that the remaining spins are in a state with definite $\tau^z$, all interactions \emph{except} those between spin I and region II are decoupled by this protocol; thus, the decay of the DEER response directly measures the influence of region II on spin I. 

Before analyzing the DEER response, we summarize our qualitative expectations [Fig.~\ref{fig:schematic}(d)]. In the diffusive phase, both spin-echo and DEER responses should decay on a fast timescale set by the spin-spin interaction. In the NIL phase, both spin-echo and DEER responses should saturate at  the same nonzero value in the thermodynamic limit, as dephasing is absent. Finally, in the MBL phase, the spin-echo response should saturate while the DEER response exhibits slow decay.

\emph{DEER response.---}The time-evolution of the many-body wave function under the DEER sequence is described by
\beq\label{deer}
| \psi(t) \rangle = R^{\pi/2}_\text{I} e^{- i {\hat H} \frac{t}{2}} R^\pi_\text{I} R^{\pi/2}_\text{II}  e^{- i {\hat H} \frac{t}{2}}R^{\pi/2}_\text{I} \ket{\psi(0)},
\eeq
where $R^{\pi/2}_r =  \prod_{j\in r}(\hat{\mathds{1}} - i \hat \sigma_j^y)/\sqrt{2}$, and $R^{\pi}_r=(R^{\pi/2}_r)^2$. 

Many features of the DEER response can be understood by keeping only two-spin interactions in Eq.~(\ref{simpleH}), in which case the response takes a compact form:
\begin{equation} \label{deerexact}
\mathcal{D}(t)  \equiv  \langle \psi(t) | \hat \tau^z_\text{I} | \psi(t) \rangle 
=   \text{Re}\prod_{j \in \mathrm{II}}  \left( \frac{1 + e^{2i \mathcal{J}_{{\rm I}j} \tau_j t}}{2} \right) 
\end{equation}
where the product is over the $N$ spins of region II and $\tau_j$ is the initial configuration of spin $j$. The additional effects induced by three- and higher-spin interactions are considered below; see also the Supplemental Material~\cite{suppmat}. 

To analyze the behavior of $\mathcal{D}(t)$, we note that the couplings $\mathcal{J}_{{\rm I}j}$ decay exponentially with the separation $|j-{\rm I}|$, and therefore different terms on the r.h.s.~of Eq.~\eqref{deerexact} oscillate at very different frequencies. This leads to a separation of scales: at a given time, there are $\sim N_{\text{fast}}$  ``fast'' coupling constants, for which $\mathcal{J}_{{\rm I}j} t \gg 1$, and the remaining ones are ``slow,'' $\mathcal{J}_{{\rm I}j} t \ll 1$. In {the product in} Eq.~(\ref{deerexact}),  the terms corresponding to slow couplings {contribute factors which are} close to $1$ and are nearly time-independent, while the terms corresponding to fast couplings oscillate between 0 and 1. Thus, $\mathcal{D}(t)$ can be separated into a time-averaged term $\bar{\mathcal{D}}(t)$ and an oscillatory term, $\mathcal{D}_{\mathrm{osc}}(t)$:
\beq\label{eq:D_structure}
\mathcal{D}(t)=\bar{\mathcal{D}}(t)+\mathcal{D}_{\rm osc}(t), 
\qquad
\bar{\mathcal{D}}(t)=1/2^{N_\text{fast}(t)},
\eeq
where the first term is obtained by replacing rapidly oscillating terms with their average value of $1/2$.

The number of ``fast'' couplings depends on time, and can be estimated knowing that $\mathcal{J}_{{\rm I}j} \propto \exp(-|j-{\rm I}|/\xi)$.  A coupling becomes ``fast''  when  $|j - I| \alt \xi \log(t)$, i.e., when entanglement has had time to propagate between the two regions~\cite{serbyn_universal_2013}. Thus, the DEER response has three regimes: (i)~at short times $t \alt t_0 \equiv \hbar/\mathcal{J}_{Ik}$ (where $k=I+d$ is the spin in region II that is most strongly coupled to I), $N_{\mathrm{fast}} = 0$ and dephasing is absent; (ii)~at intermediate times $t_0 \alt t \alt t_0 e^{N/\xi}$, we find $N_{\text{fast}}(t)\sim \xi \log (t/t_0)$, so that $\bar{\mathcal{D}}(t) \sim t^{- \xi \ln 2}$; and (iii)~at very long times $t \gg t_0 e^{N/\xi}$, all couplings are fast, so that the DEER response saturates at $\bar{\mathcal{D}}(\infty) \approx 2^{-N}$. These three regimes can be combined using the following interpolation formula: 
\beq\label{eq:D_time}
\bar{\mathcal{D}}(t) = \left\lbrace \begin{array}{cl} {( 1+t^2/t_0^2 )^{-\alpha/2}} & \quad t \lesssim t_0 e^{N/\xi} \\ 2^{-N} & \quad t \gg t_0 e^{N/\xi} \end{array} \right. ,
\eeq
where $\alpha=\xi \ln 2$. Upon disorder averaging, one expects $\mathcal{D}_{\mathrm{osc}}(t)$ to be suppressed, as the oscillation frequencies vary randomly from realization to realization. Thus the full disorder-averaged DEER response is given by \eq{eq:D_time}.

We note that, although truncating Eq.~\eqref{simpleH} at two-spin interactions gives the correct structure for the time- and disorder-averaged DEER response, it leads to incorrect predictions for the oscillatory term $\mathcal{D}_{\mathrm{osc}}(t)$. Three- and higher-spin terms make the oscillation frequencies dependent on the initial eigenstate, leading to the suppression of $\mathcal{D}_{\mathrm{osc}}(t)$ upon thermal averaging (see~\cite{suppmat} for details). 

 \begin{figure}[t]
\begin{center}
\includegraphics[width=0.45\textwidth]{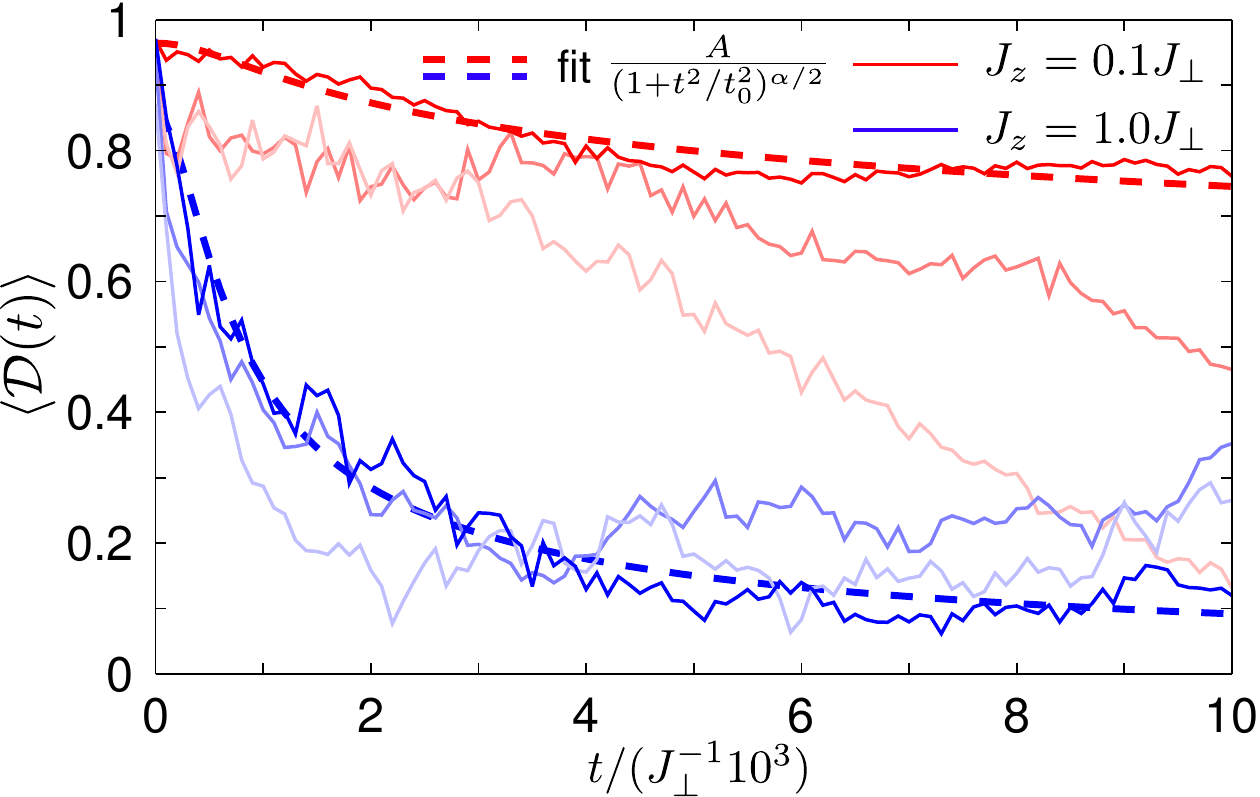}
\caption{DEER response thermally averaged over 50 eigenstates for three particular disorder realizations of the random-field XXZ model, both at weak interactions ($J_z = 0.1 J_\perp$) and moderate interactions ($J_z = J_\perp$). The general trend is consistent with that predicted by Eq.~\eqref{eq:D_time}, but residual oscillations and sample-to-sample fluctuations are strong. The disorder strength is $W=6J_\perp$; spin I is located at $\text{I}=3$, and separated by $d=3$ spins from region II with $N=7$ spins. 
}
\label{fig3}
\end{center}
\end{figure}

\begin{figure}[t]
\begin{center}
\includegraphics[width=\columnwidth]{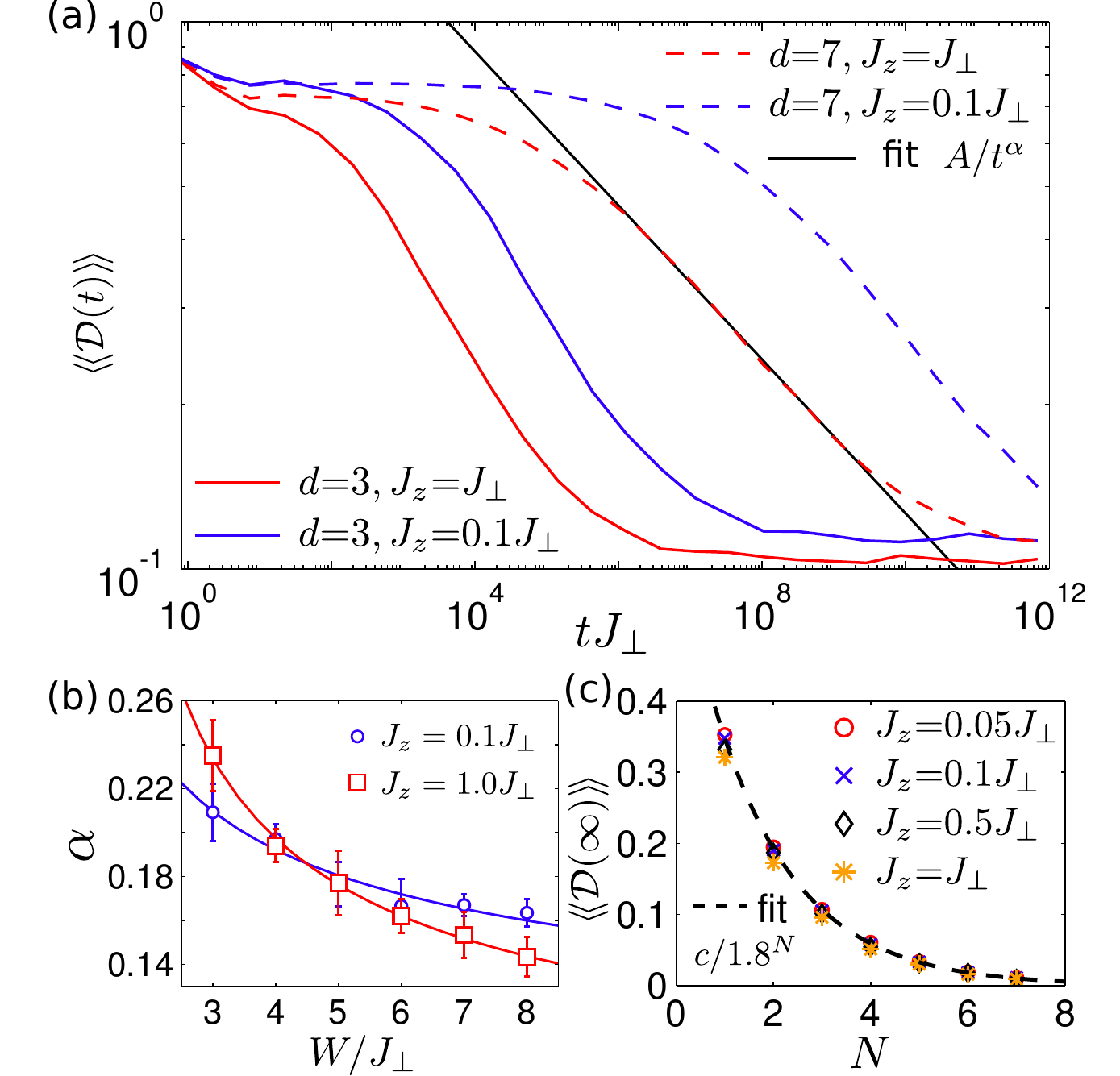}
\caption{  (a) Disorder-averaged DEER response for the random-field XXZ model for both weak ($J_z = 0.1 J_\perp$) and moderate ($J_z = J_\perp$) interactions, and both small ($d = 3$) and large ($d = 7$) separations between spin I and region II. A regime of power-law decay, spanning multiple decades, can be seen in all cases. (The disorder strength is $W = 8 J_\perp$, and the size of region II is $N = 3$.) (b) Dependence of the exponent $\alpha$ on disorder $W$ (for $d=7$ and $N=3$). 
The fit, given by $\alpha=c_1/\ln (c_2 W)$, is consistent with the scaling of the localization length, $\xi \sim 1/\ln(W)$, at strong disorder. (c) Saturated value of DEER response, $\mathcal{D}(\infty)$, for various sizes of region II (denoted $N$) and various values of $J_z$. We find that $\langle\langle\mathcal{D}(\infty)\rangle\rangle$ does not depend on $J_z$ and decreases with $N$ as $c/1.8^N$.}
\label{fig4}
\end{center}
\end{figure}

{\it Numerical simulations.}---We now test the previous arguments against numerical simulations by studying the spin-echo fidelity and DEER response for a 1D random-field XXZ spin chain, believed to exhibit an MBL phase~\cite{pal_many-body_2010}: 
\begin{equation}
 \hat H = \frac{J_\perp}{2} \sum_{\langle ij \rangle}  (\hat S_i^+ \hat S_{j}^- +  \hat S_{j}^+ \hat S_{i}^-) 
 + J_z \sum_{\langle ij \rangle} \hat S_i^z \hat S_{j}^z +  \sum_i h_i  \hat S_i^z\label{eq:h}
\end{equation}
 where $\hat{\mathbf S}_j$  are spin-1/2 operators with eigenvalues $\pm 1/2$, $\hat S^\pm_j = \hat{S}_j^x \pm i \hat S_j^y$, and the random field $h_i$ is uniformly distributed in the interval $ [-W;W]$. For open boundary conditions and $J_z=0$, $\hat H$ maps onto free fermions moving in a disorder potential. In this limit, the system is in an NIL phase for any $W > 0$. When $J_z \neq 0$, the system is believed to exhibit both MBL and delocalized phases as a function of $W/J_\perp$~\cite{pal_many-body_2010}. 

Although the Hamiltonian in the MBL phase can be expressed in the form of Eq.~\eqref{simpleH} when written in the basis of conserved quantities (effective spins), in experiments one manipulates the physical $S$-spins, rather than the effective $\tau$-spins. In what follows, we calculate the response for spin-echo and DEER protocols performed on the physical spins. We show that, due to the local relation between physical and effective spin operators, the behavior of spin-echo and DEER responses discussed above remains qualitatively correct throughout the MBL phase, and becomes quantitatively correct in the limit of strong disorder. 

We study time evolution and response functions by exact diagonalization of the Hamiltonian (\ref{eq:h}). Unless otherwise specified, the chain contains $L=12$ spins with open boundary conditions. The Hamiltonian is diagonalized for all total $S^z$ sectors, and DEER response $\mathcal{D}(t)  \equiv  \langle \psi(t) | \hat \tau^z_\text{I} | \psi(t) \rangle$ is calculated numerically. The initial state $ | \psi(0) \rangle $ is a randomly chosen eigenstate, such that ${\cal D}(0) >0$. 
Thermal averaging is performed over the entire band (infinite temperature), and is denoted by single brackets, $\langle {\cal D}(t) \rangle $. 

We first calculate thermally-averaged spin-echo and DEER response for a single disorder realization (Fig.~\ref{fig2})~\cite{fnote1}. In the MBL phase, the spin-echo fidelity remains finite at long times, but its saturation value is smaller than one,  reflecting the difference between physical and effective spins. Each pulse affects several effective spins; therefore, the probability to come back to the initial state at the end of the sequence is reduced. At strong disorder, the spin-echo fidelity saturates to a value close to unity and is system-size independent. In contrast, the DEER response decays to values much smaller than one.  

Fig.~\ref{fig3} demonstrates that the DEER response (thermally averaged over 50 eigenstates for a single disorder realization) fits well to the modified interpolation formula $\mathcal{D}(t) = A/(1+t^2/t_0^2)^{\alpha/2}$, where a multiplier $A$ has been introduced to account for the difference between effective and physical spins [cf. \eq{eq:D_time}]. However, the oscillations coming from $\mathcal{D}_{\rm osc}(t)$ are still significant.

Plotting the disorder-averaged DEER response (denoted by double brackets $\llangle \mathcal{D}(t) \rrangle$) on a double logarithmic scale, Fig.~\ref{fig4}(a), clearly reveals a power law decay spanning several decades. Comparison of $\llangle \mathcal{D}(t) \rrangle$ for different separations $d$ between regions I and II illustrates the sensitivity of $t_0 [\sim \exp (d/\xi)]$ to $d$. Fig.~\ref{fig4}(b) shows the dependence of the exponent governing the power-law decay, $\alpha$, on disorder: $\alpha$ decreases with increasing disorder strength, and is well-described by the functional form $\alpha=c_1/\ln (c_2 W)$, 
consistent with the relation $\alpha=\xi \ln 2$ and scaling of the localization length $\xi \propto 1/\ln (W)$ at strong disorder. %
Finally, we study the dependence of the disorder-averaged saturation value of the DEER response as a function of the number of spins $N$ in region II, Fig.~\ref{fig4}(c). The saturation value, which is nearly independent of the interaction strength, fits to a function $f(k) = c/1.8^N$ (for effective spins, by contrast, Eq.~\eqref{eq:D_time} predicts $1/2^N$). Thus, the DEER response for physical spins has the same functional form as that for effective spins, although the coefficients are different, owing to the difference between physical and effective spin operators. 

\emph{Experimental considerations.---}Promising experimental systems for studying MBL include ultracold atomic~\cite{demarco1, demarco2, Inguscio} and molecular~\cite{polar,yan_observation_2013} gases confined in optical lattices, as well as localized spin defects in solids such as nitrogen-vacancy (NV) centers in diamond~\cite{NV1,NV2}. Such systems can be well-isolated from their environment 
and feature long coherence times. Further, they can be manipulated by optical and microwave fields, thus 
allowing for implementation of the pulsed protocols. 
We now evaluate the feasibility of the present protocols in a number of experimental settings. In each case, the slow DEER decay can be observed provided that: (a) there exists a separation of scales between the couplings $J_\perp, J_z$ and the extrinsic decoherence rate $T_1^{-1}$, and (b) excitations are localized on a small number of sites, ensuring a reasonable spin-echo fidelity. 

The most direct implementation of our protocols involves a two-component Fermi- or Bose-Hubbard model in a disordered optical lattice: in such models, random spin-spin interactions arise via superexchange, and random fields can be imposed via a state-dependent optical lattice. The typical interaction scale $J \approx 10$ Hz, whereas achievable $\text{T}_1$ times limited by particle loss are about $10$s~\cite{strohmaier_observation_2010,trotzky_time-resolved_2008,fukuhara_quantum_2013}. The ratio between these scales is $\alt 500$; thus, the DEER protocol can detect entanglement at realistic distances $\alt \xi \ln(J \text{T}_1) \approx 6 \xi$. 
Even more favorable conditions exist in systems with dipolar interactions. For instance, in NV-center samples at achievable densities 
(e.g., spacings of $10$ nm), 
$J \sim 100$ kHz and $\text{T}_1\sim10$ ms, yielding $\text{T}_1/J^{-1} \sim 5 \times 10^3$.
For Rydberg atoms, $J\sim(1-10)$MHz (e.g. in 38s state of Rb at typical distances $\approx 5 \mu$m), while $\text{T}_1\sim 100\mu$s; therefore, $\text{T}_1/J^{-1} \sim (0.5-5) \times 10^3$. 
Finally, for polar molecules in optical lattices, $J\sim 50$Hz and $\text{T}_1\sim25$s,
and thus $\text{T}_1/J^{-1} \sim 8 \times 10^3$. 
For all these cases, therefore, it should be feasible to probe interaction effects 
in the MBL phase through DEER; however, the functional form of the dephasing might differ from that considered here, as  the power-law  tails of the dipolar interactions affect localization (although the MBL phase is expected to survive for dipolar interactions in one dimension~\cite{levitov1990,burin, yao13}). 

Before concluding, we note that since the proposed protocols can distinguish various phases  \emph{after} disorder-averaging, they can be applied even in experiments where the disorder realization changes between 
individual experimental runs. This is especially important for realizations involving atoms or molecules loaded at random into a deep optical lattice; in such systems each disorder-realization is destroyed upon measurement.

\textit{Summary.---}In summary, we showed that coherent manipulation of spins can be used to probe many-body localization. In particular, the modified spin-echo protocol directly probes the characteristic slow entanglement growth in the MBL phase, and distinguishes it from the NIL phase and the delocalized phase. We demonstrated that the corresponding response function exhibits a power-law time decay, which reflects the broad distribution of time scales present in the MBL phase. The technique is robust with respect to thermal and disorder averaging, and can be implemented, using currently accessible experimental means, in ultracold atomic, molecular and solid-state spin systems. 

\textit{Acknowledgements.} We thank E. Altman, Y. Bahri, I. Bloch, T. Giamarchi, D. Huse, V. Oganesyan, A. Pal, D. Pekker, and G. Refael for insightful discussions.The authors acknowledge support from the Harvard Quantum Optics Center, Harvard-MIT CUA,  DARPA OLE program, AFOSR Quantum Simulation MURI, ARO-MURI on Atomtronics, ARO-MURI Quism program, and the Austrian Science Fund (FWF) Project No. J 3361-N20. Simulations presented in this article were performed on computational resources supported by the High Performance Computing Center (PICSciE) at Princeton University and the Research Computing Center at Harvard University.

\newpage

\onecolumngrid
\appendix

\begin{center}
{\large \bf Supplemental Material for\\ Interferometric signatures of many-body localization}
\end{center}

\subsection{Higher-order spin-spin interaction terms}

We  analytically calculate the full DEER response in the MBL phase which is effectively described 
by the Hamiltonian \eqw{simpleH}, assuming that one directly manipulates the effective spins. We will show that at long times the full DEER response is described by Eq.(\ref{eq:D_structure}) of the main text. It will also become evident that the oscillatory term $\mathcal{D}_{\rm osc}$ behaves differently when higher-spin interactions are taken into account, compared to the approximation when only two-spin interactions are kept.

The DEER response for a system initially prepared in a product state $|\tau_1 \tau_2 \tau_3 \ldots \rangle$ is given by
\begin{equation}
\label{eq:deerGen}
\mathcal{D}(t) = \re \frac{1}{2^N} \sum_{\lbrace \upsilon \rbrace \in 2^N} \prod_{j\in\text{II}} e^{i \mathcal{\tilde J}_{Ij} (\tau_j+\upsilon_j)t} \prod_{j\neq i\in\text{II}} e^{i \mathcal{\tilde J}_{Iji} (\tau_j \tau_i -{\upsilon_j \upsilon_i})t}\prod_{j\neq i\neq l\in\text{II}}\ldots
\end{equation}
where $N$ is the size of region II, $\sum_{\lbrace \upsilon \rbrace\in 2^N}$ is a sum over all 
$2^N$ spin configurations in region II, and $\mathcal{\tilde J}_{Ij} = \mathcal{J}_{Ij} + 
\sum\nolimits_p \mathcal{J}_{Ijp} \tau_p + \ldots$ are the couplings that are ``dressed'' by  
the spins $p$ outside regions I and II. For vanishing 
three-body $\mathcal{J}_{Ijk}=0$ and higher-order $\mathcal{J}_{Ijk\ldots}=0$ interactions, we obtain Eq.~(3) 
of the main text as the sum over the spin configurations in region II can be directly evaluated.
\begin{figure}[b]
\begin{center}
\includegraphics[width=.65\columnwidth]{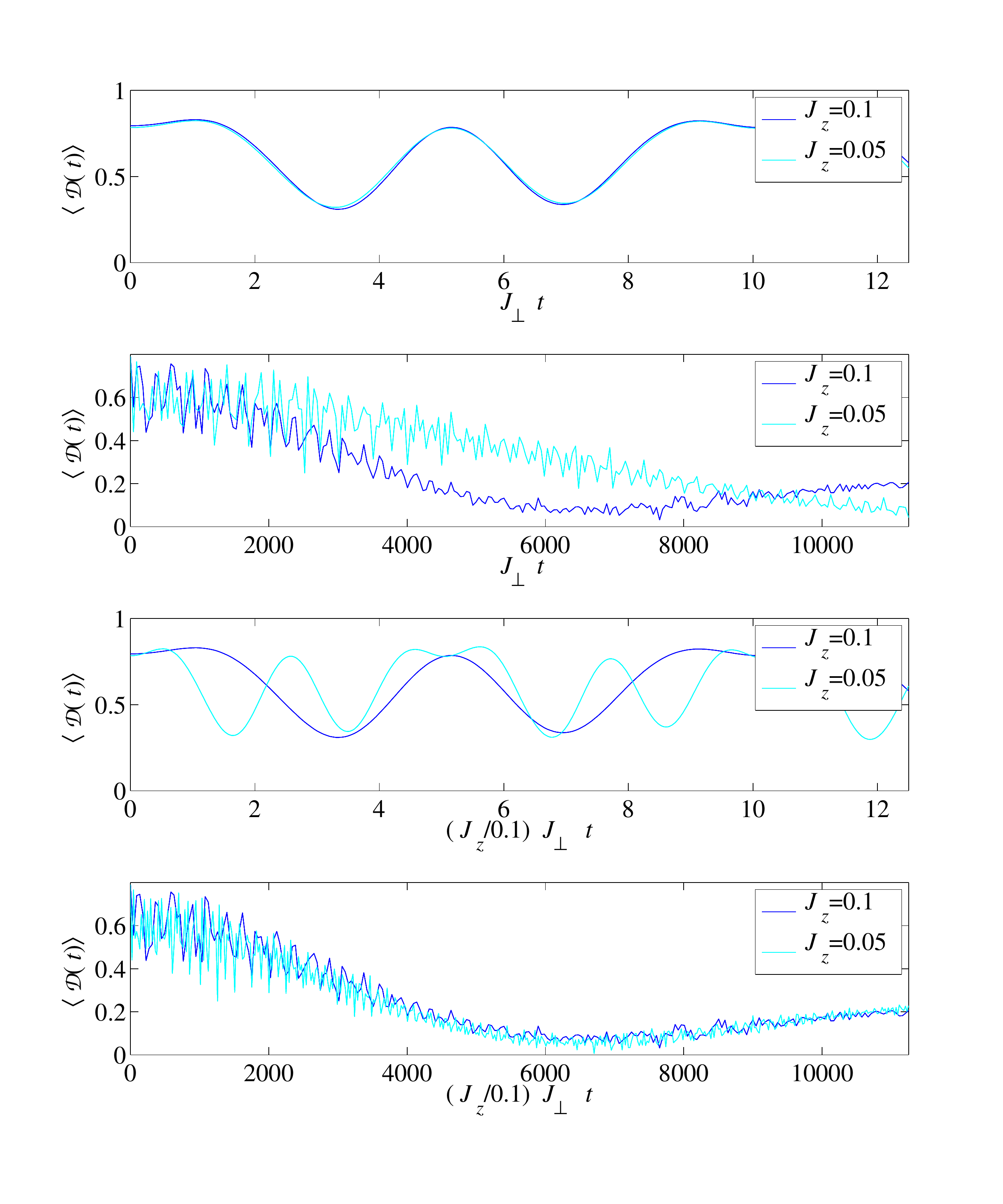}
\caption{ DEER response for a single disorder realization averaged over 500 eigenstates. Different curves correspond to the different interaction strength $J_z$. Top two plots show DEER signal for two different time scales. (a) On short timescales the dephasing is irrelevant and curves collapse. For longer times, panel (b), the interaction-induced dephasing becomes relevant and signal depends on the interaction strength. Panels (c) and (d) show the same data plotted as a function of $J_z t$. In the last panel curves collapse, proving that this dynamics is governed primarily by the dephasing coming from two-body interaction terms. Note that on even longer time-scales, the three-body (and higher) interaction terms become relevant and signal differs even when plotted as a function of $J_z t$ (not shown). Disorder $W=5J_\perp$, spin I is located at $\text{I}=3$, and is separated by $d=3$ spins from region II with $N=4$ spins.
}
\label{figSupp2}
\end{center}
\end{figure}

We now argue that Eq.(\ref{eq:deerGen}) gives rise to the time-averaged DEER response of the form (\ref{eq:D_structure}), previously obtained by keeping only two-spin terms in Eq.(\ref{simpleH}). The couplings $\mathcal{J}_{{\rm I}j}, \mathcal{J}_{{\rm I}ji}, ...$ decay exponentially with the distance between spin {\rm I} and spins $j,i$, etc., and therefore a scale separation arises: at a given time, there are fast couplings, for which $\mathcal{J}_{{\rm I}ji}t\gtrsim 1$, and slow ones,  $\mathcal{J}_{{\rm I}ji}t\ll 1$. At $t\lesssim t_0$, all couplings are slow, and there is no dephasing. At $t_0\lesssim t \lesssim t_0 e^{N/\xi}$, couplings which involve spins $k,k+1,..k+N_{\rm fast}(t)-1$, where $N_{\rm fast}(t)\sim \xi \log t/t_0$, are fast, while the remaining ones are slow. Thus, $\mathcal D(t)$ is a sum of $2^{N_{\rm fast}(t)}$ terms, one of which is constant and given by $1/2^{N_{\rm fast}}$, while the remaining ones are rapidly oscillating. Thus, time-averaged DEER response is given by $\bar{\mathcal{D}}(t)\approx (t_0/t)^\alpha$, $\alpha=\xi \ln2$. Finally, at $t\gtrsim t_0 e^{N/\xi}$ all couplings are fast, and the time-averaged response saturates at a value $\bar{\mathcal D}(\infty)\approx 2^{-N}$. Thus, including higher-spin terms leads to the same behavior of the time-averaged DEER response as we obtained in the main text. 

However, the oscillatory term behaves qualitatively differently upon inclusion of higher-order spin-spin interaction terms. Performing the thermal average entails averaging the DEER response over all $2^N$ configurations of the $\{ \tau_p \}$. When higher-order terms are included, for a given initial configuration, the oscillatory term consists of $2^N-1$ terms oscillating with independent frequencies. Moreover, the frequencies also depend on the state of the spins situated between regions I and II. Therefore, we expect that the oscillations would be strongly suppressed by thermal averaging. 

In contrast, when only two-spin interactions are taken into account, there are only $N$ independent frequencies set by $\mathcal{J}_{{\rm Ij}}$, $j=k,...k+N-1$. Therefore, we expect the oscillations to remain much stronger, and not to be washed out by thermal averaging.

\subsection{Limit of weak interactions}

We note that in realistic models, such as the random-field XXZ model, the higher-order interactions between effective spins generally will appear in higher orders in perturbation theory in interaction strength $V$~\cite{serbyn_universal_2013}. For example, three-spin terms are proportional to $V^2$. Therefore, at very weak interactions, there will be an intermediate time regime, in which higher-spin terms are not effective in suppressing oscillations around the averaged DEER response. Since in this regime only two-spin couplings, which are proportional to $V$, give rise to dephasing, all response functions are expected to depend only on the product $Vt$. We illustrate this behaviour in Fig.~\ref{figSupp2}.

\subsection{DEER protocol for effective model of MBL phase}

We illustrate the DEER response for the {\it effective} model \eqw{simpleH}, in which we assume the following distribution of coupling constants:
\begin{equation}
 \mathcal{J}_{ij}=r_{ij} \exp\left(-|i-j|/\xi\right), \, \mathcal{J}_{ijk}=r_{ijk} \exp\left(-[|i-j|+|j-k|+|k-i|]/2\xi\right),\, \ldots,
 \label{effCoupl}
\end{equation}
where $\xi$ is the localization length and the amplitudes $r_{ij\ldots}$ are \emph{independent} random variables drawn from box disorder of relative 
width $\mathcal{W}$ which is centered around $1$. (In realistic models such as the XXZ model, by contrast, the various couplings should in general be correlated.) 

\begin{figure}[t]
\begin{center}
\includegraphics[width=.8\columnwidth]{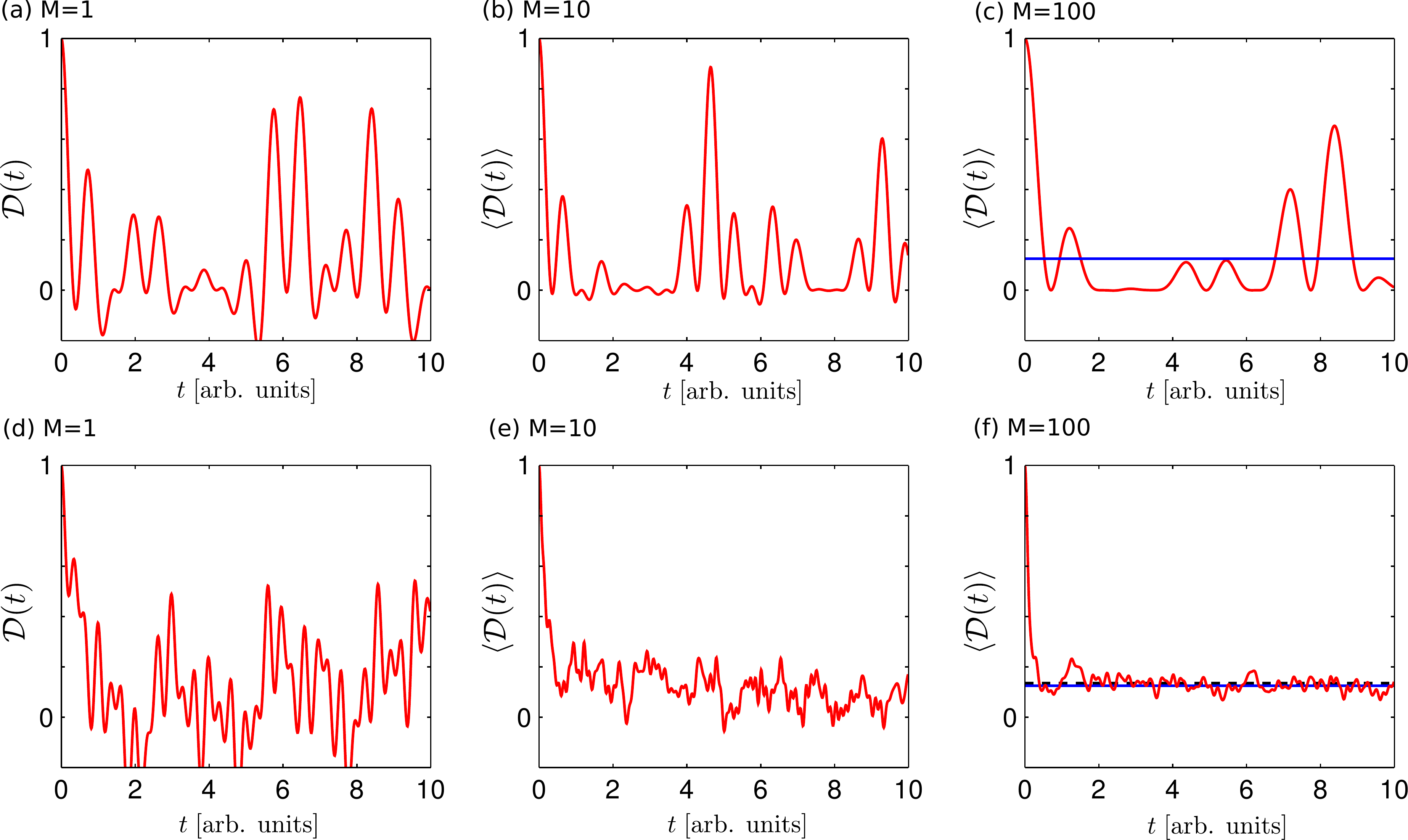}
\caption{ DEER response \eqw{eq:deerGen} evaluated for the effective Hamiltionian \eqw{simpleH} using
coupling parameters $\mathcal{J}_{ij\ldots}$ given by \eq{effCoupl}. The top row shows the DEER response
for two-body interactions only while the bottom row shows the response including two- and three-body 
interactions. The DEER response is averaged over $M$ intitial spin configurations. The localization length is set to $\xi=1$ and are width of 
the box disorder is $\mathcal{W}=20\%$. Higher-order spin interactions suppress the oscillations in the DEER response.
}
\label{figSupp1}
\end{center}
\end{figure}

The DEER response \eqw{eq:deerGen} for this model is shown in \fig{figSupp1}.  The upper row shows the response when only two-body interactions are present, while bottom row illustrates the response when both two-body and three-body interactions are present. The response is averaged over $M$ initial spin configurations $|\tau_1 \tau_2 \tau_3 \ldots \rangle$. The results demonstrate that the dressing of the couplings by spins outside regions I and II, which is generated by higher-order spin interactions, suppress the oscillations in the DEER response, in agreement with arguments given in the previous section. Including four- and higher-spin terms does not qualitatively change the behavior of the DEER response compared to the case when only three-spin interactions are included.

\begin{thebibliography}{100}
\bibitem{fnote1}%
  \BibitemOpen
  \bibinfo {note} {{Spin echo was implemented both using the protocol of Fig.~\ref{fig:schematic}(b) and a modified protocol in which a $\pi/2$ pulse was applied to region II at time $t = 0$; the disorder-averaged fidelity shows the same behavior for both protocols.}}%
  \bibAnnoteFile{NoStop}{fnote1}%
\end{thebibliography}
\end{document}